\documentstyle[12pt]{article}
\newcommand{\rf}[1]{(\ref{#1})}
\newcommand{\beq}{\begin{equation}}
\newcommand{\eeq}{\end{equation}}
\newcommand{\bdm}{\begin{displaymath}}
\newcommand{\edm}{\end{displaymath}}
\newcommand{\bea}{\begin{eqnarray}}
\newcommand{\eea}{\end{eqnarray}}
\newcommand{\nn}{\nonumber \\}

\newcommand{\dg}{\dagger}
\newcommand{\g}{\gamma}

\newcommand{\Del}{\Delta}

\newcommand{\ve}{\varepsilon}

\newcommand{\lam}{\lambda}
\newcommand{\mn}{\mu\nu}

\newcommand{\pa}{\partial}

\newcommand{\th}{\theta}

\newcommand{\ra}{\rightarrow}

\newcommand{\ws}{Weinberg--Salam}
\newcommand{\ym}{Yang--Mills~}
\newcommand{\lag}{Lagrangian~}

\def\gtwid{\raise.3ex\hbox{$>$\kern-.75em\lower1ex\hbox{$\sim$}}}
\def\ltwid{\raise.3ex\hbox{$<$\kern-.75em\lower1ex\hbox{$\sim$}}}
\begin{document}
\topmargin -0.8cm
\headheight 0pt
\headsep 0pt
\topskip 9mm

\begin{flushright}
NBI-HE-93-58\\
\hfill
October 1993
\end{flushright}

\hspace{3.5cm}

\begin{center}
{\Large \bf A W--DRESSED ELECTROWEAK STRING}
\end{center}

\vspace{30pt}

\begin{center}
{\large \bf P. Olesen}

\vspace{30pt}

The Niels Bohr Institute \\
University of Copenhagen \\
Blegdamsvej 17 \\
DK--2100 Copenhagen \\
Denmark
\end{center}

\vspace{3.5cm}
\begin{center}

{\bf Abstract}
\end{center}

We give plausibility arguments for the existence of a $W$--dressed electroweak
string for the \ws ~model with $\th_W = 0$.
This string is a $Z$--string which in the core has a finite energy
contribution from $W$--condensation induced by the anomalous magnetic moment
in the \ym field.
The solution which has minimum energy at $r=0$ interpolates between
the unbroken $(r=0)$ and the broken $(r \ra \infty) \ SU(2) \times
U_y(1)$ phase.

\newpage
Recently it has been shown by Vachaspati \cite{1} that the electroweak \ws
\ theory \cite{2} has vortex solutions.
This solution is formed from the lower component of the Higgs field
and the $Z$ field, and it satisfies the usual vortex equations of motion
\cite{3}.
These electroweak strings are not stable on topological grounds,
and hence their stability must be investigated.
This was done \cite{4}, and stability was found for a range of
parameters unfortunately not within the experimental limits.
It has been suggested \cite{5} that perhaps the inclusion of at least
two Higgs multiplets may improve this situation.

It has, however, been pointed out by Perkins \cite{6} that the
electroweak string can have its energy lowered by the formation of a
$W$--condensate in its core\footnote{This point was also mentioned
by the author in a communication to T. Vachaspati.}.
This is because of the fact that the electroweak vacuum is unstable
to the formation of a $W$--condensate \cite{7} if a strong ``magnetic''
$Z$--field is applied.
The physical reason for this effect is that the electroweak vacuum is
``paramagnetic'', with an energy which contains the terms
\beq
- \cos \th_W {\bf B}^Z~\mbox{\boldmath $\mu$} + \mbox{\boldmath $\mu$}^2~~~~,
                               \label{1}
\eeq
where $B^Z_i = \frac{1}{2} \ve_{ijk} Z_{jk}$ and the magnetic
moment density \mbox{\boldmath $\mu$} is given by\\
$\left( W_i = \sqrt{\frac{1}{2}} (W^1_i + iW^2_i)\right)$
\beq
\mu_j = ig \ve_{jkl} W^\dg_k W_l \label{2}
\eeq
Although the energy contains many other terms
than those exhibited in eq. \rf{1} it is clear that
eq. \rf{1} can give rise to a $W$--condensation under some circumstances.

The result obtained by Perkins \cite{6} shows that it is possible to
find $W$--configurations which lower the energy of the $Z$--string,
provided $\sin \th_W < 0.9$.
In this connection it should be noticed that the first term in
eq. \rf{1} vanishes for $\th_W \ra \pi /2$, so for $\th_W$ close
to $\pi /2$ very little energy is gained by $W$--condensation.
Thus, for the experimental value $\sin^2 \th_W \approx 0.23$ ~
$W$--condensation should occur in the core of the $Z$--string.

In this paper we present plausability arguments that there exists a
``$W$--dressed $Z$--string'' for small values of $\th_W$.
In order to simplify the calculations we take $\th_W = 0$ in the
following.
This should be satisfactory as far as a first orientation is
concerned, since the usual $Z$--string is stable near $\th_w = \pi /2$
\cite{4}, and hence the problem of instability is most marked near
$\th_W = 0$.

In order to have a solution for a $W$--dressed electroweak string,
we need to satisfy all the equations of motion of the electroweak
string.
In order to produce an ansatz which can work, one has to be very
careful.
Let us start with the Higgs field $\phi$.
We would like to have only a single complex field.
Let us therefore start by asking if the upper complex component
$\phi_U$ in
\beq
\phi = \left( \begin{array}{l} \phi_U \\
\phi_L   \end{array} \right)                       \label{3}
\eeq
can be put to zero.
This is not a trivial problem:\footnote{I thank T. Vachaspati for
pointing this out to me.}
Consider the equation of motion for $\phi_U$ obtained from the \lag
by varying $\phi^\ast_U$.
This equation has the following structure:
\bea
(\mbox{terms which vanish when}~~\phi_U  = 0)
+ \ \frac{1}{2} ig \left(W^\dg_i \pa_i \phi_L + \pa_j
(W^\dg_ j \phi_L)\right)  =  0~~~~.
                                                          \label{4}
\eea
The term written out explicitly apparently does not vanish for
$\phi_U = 0$.
If this is correct, then $\phi_U = 0$ is not a solution of the
equations of motion \rf{4}.

However, it turns out by closer inspection that the explicit term
in eq. \rf{4} does actually  vanish for $\phi_U = 0$.
This can be seen from the equations of motion for $W_\mu$ and
$Z_\mu$.
These equations together produce an integrability condition.
In order to simplify the equations, we start by taking
$\phi_U = 0~,~ \phi_L \neq 0$.
It then turns out that the explicit term in eq. \rf{4} vanishes
because of the integrability condition, and hence it is perfectly
consistent to have $\phi_U= 0$.

For the moment, consider an arbitrary $\th_W$.
The equation for $W_\mu$ is then
\bea
D_\mu (D_\mu W_\nu - D_\nu W_\mu)& - &ig F^3_{\mn} W_\mu - \frac{1}{2} g^2
| \phi_L |^2 W_\nu = 0~~~~,   \label{5}    \\
F^3_{\mn}  = \pa_\mu V^3_\nu & - &\pa_\nu V^3_\mu - ig(W^\dg_\mu W_\nu
- W^\dg_\nu W_\mu)~~~~, \label{6}        \\
V^3_\mu =  Z_\mu \cos \th_W &  + & A_\mu \sin \th_W~~~,~~~D_\mu =
\pa_\mu + ig V^3_\mu~~~~.         \label{7}
\eea
Now operate with $D_\nu$ on eq. \rf{5}, which then reduces to
\beq
- g^2(W^\dg_\mu W_\nu - W^\dg_\nu W_\mu) D_\nu W_\mu + ig~\pa_\nu F^3_{\mn}
W_\mu + \frac{1}{2}~ g^2 D_\nu ( | \phi_L|^2 W_\nu) = 0   ~~~~.\label{8}
\eeq
In this expression we can use the equation for $F^3_{\mn}$,
\beq
- \pa_\mu F^3_{\mn} + ig(W_\mu F^\dg_{\mn} - W^\dg_\mu F_{\mn})
+ \frac{g^2}{2 \cos \th_W} |\phi_L|^2 Z_\nu
+ \frac{ig}{2} (\phi_L \pa_\nu \phi^\ast_L - \phi^\ast_L \pa_\nu \phi_L) =
0~~~~.
                                                           \label{9}
\eeq
Using this in eq. \rf{8} we finally get
\beq
D_\nu (W_\nu |\phi_L|^2 ) - \frac{ig}{\cos \th_W} |\phi_L|^2 Z_\mu W_\mu
+ (\phi_L \pa_\mu \phi^\ast_L - \phi_L^\ast \pa_\mu \phi_L) W_\mu = 0~~~~.
                                          \label{10}
\eeq
A similar expression was first derived by MacDowell and Törnkvist
\cite{8} for the case where $\phi_L$ is real.
For complex $\phi_L$ one gets the last term in eq. \rf{10} as an
additional term.

For $\th_W = 0$ eq. \rf{10} reduces to
\beq
\pa_\nu (W_\nu |\phi_L|^2 ) + (\phi_L \pa_\mu \phi^\ast_L - \phi^\ast_L \pa_\mu
\phi_L) W_\mu = 0~~~~.                               \label{11}
\eeq
Now let us return to the original problem in eq. \rf{4}, which we can
rewrite ($\chi$ is the phase of $\phi_L$)
\bea
\lefteqn{(\mbox{terms which vanish when}~~ \phi_U = 0) \ + }\nn
& &+ \frac{1}{2} ig \frac{e^{i \chi}}{|\phi_L|} \left[ \pa_i
(W^\dg_i |\phi_L|^2 ) + (\phi^\ast_L \pa_i \phi_L - \phi_L \pa_i
\phi^\ast_L ) W^\dg_i \right]  = 0~~~~.                \label{12}
\eea
The quantity in the square bracket vanishes because of eq. \rf{11}.
Thus we can take $\phi_U = 0$ because this is consistent with the equation
of motion for $\phi_U$.
In the following $\phi_L$ is denoted $\phi$, and the Higgs field has the
form $\left( \begin{array}{c} 0 \\ \phi \end{array} \right)$.

Since we are interested in a $W$--dressed $Z$--string we take the
$Z$--field to have the usual structure $Z_{r \th} = - Z_{\th r} \neq
0~~,~~Z_\th \neq 0$, and all other $Z$ fields equal to zero
\cite{1,3}.
Also, from the integrability condition \rf{11} it is easy to see that
in order to have a non--trivial solution we need two non--vanishing
$W$--fields.
Since $Z_{r \th}$ couples to $W_\th$ and $W_r$ (but not $W_z$ and $W_0$)
we take $W_\th$ and $W_r$ to be non--vanishing, and $W_z = W_0 = 0$.
The $W$--equation \rf{5} then gives
\bea
\frac{1}{r}\frac{\pa}{\pa r} r \frac{\pa W_\th}{\pa r}& - &\frac{1}{r}~
\frac{\pa^2 W_r}{\pa r \pa \th} + \frac{1}{r^2}~\frac{\pa W_r}{\pa \th} -
\frac{W_\th}{r^2}                                  \nn
& - &ig Z_{r \th} W_r -  \frac{g^2}{2} |\phi|^2 W_\th + g^2
(W_r W^\dg_\th - W_\th W^\dg_r) W_r = 0~~~~, \label{13} \\
\frac{1}{r^2}\frac{\pa^2 W_r}{\pa \th^2}& - & \frac{1}{r}~
\frac{\pa^2 W_\th}{\pa r \pa \th} + \frac{1}{r^2}~\frac{\pa W_\th}{\pa \th} -
g^2 Z^2_\th W_r     \nn
& + & ig Z_{r \th} W_\th - \frac{g^2}{2} |\phi|^2 W_r - g^2
(W_r W^\dg_\th - W_\th W^\dg_r) W_\th = 0~~~~.    \label{14}
\eea
In addition to these equations it is important that we satisfy the
integrability condition \rf{11},
\beq
\frac{1}{r} ~\frac{\pa r |\phi|^2 W_r}{\pa r}~
+ \frac{1}{r} \frac{\pa |\phi|^2 W_\th}{\pa \th} - 2i |\phi|^2 W_\th
\frac{m}{r} = 0                         \label{15}
\eeq
where we used $\phi = |\phi| e^{im \th}$, so $m$ is the winding number
for the $Z$--string.

The integration condition \rf{15} incorporates features
of the $Z$--equation \rf{9} not incorporated in eqs. \rf{13} and \rf{14}.
Any ansatz for $W_r$ and $W_\th$ is therefore in danger of being in
conflict with eq. \rf{15}, even if eqs. \rf{13} and \rf{14} are
satisfied.
Consistency of all three eqs. \rf{13} - \rf{15} is therefore a
non--trivial check of any ansatz for $W_r $ and $W_\th$.

We take the $\th$--dependence of $W_r$ and $W_\th$ to be
\beq
e^{il \th} ~~~~.                 \label{16}
\eeq
Eq. \rf{15} then gives
\beq
W_\th i(l - 2m) = - \frac{1}{|\phi|^2}~\frac{\pa r |\phi|^2 W_r}{\pa r}~~~~.
                                          \label{17}
\eeq
For a $Z$--string $|\phi| \propto r$ for $r \ra 0$.
We shall now find $W_\th$ and $W_r$ for small values of $r$.
Writing
\beq
e^{- il \th} \ W_r = ar^p + \ldots~~~~.                   \label{18}
\eeq
with $p \geq 0$ to avoid singularities, we get from eq. \rf{17} and
$|\phi| \propto r$,
\beq
e^{- il \th} \ W_\th = ia~ \frac{p + 3}{l - 2m} ~r^p + \ldots~~~~.
\label{19}
\eeq
Thus we see that because of the integrability condition the winding
number enters in $W_\th$ in the lowest order.
This is not a consequence of eqs. \rf{13} and \rf{14}.

Inserting \rf{18} and \rf{19} in eq. \rf{13}, we see that since
$Z_{r \th} \ra \mbox{const}~,~ Z_\th \ra \mbox{const} \times r$
the dominant terms are the four first terms.

Hence
\bdm
a \left[ \frac{i (p + 3)}{l - 2m} p^2 - ilp + il -
\frac{i(p+ 3)}{l - 2m} \right] r^{p-2} = 0~~~~,
\edm
or
\beq
p = - 2 + \sqrt{1 + l(l - 2m)}~~~~.     \label{20}
\eeq
In the following we take $m=1$. Then\footnote{If $m = 2,3, \ldots \
p$ is no longer an integer.
If the $W$'s are continued to complex values of $r$ a cut appears.}
\beq
p = - 2 + |l - 1|~~~~,                       \label{21}
\eeq
so $p \geq 0$ for $l = - 1, -2, \ldots$ or $l = 3,4, \ldots$~.
{}From eq. \rf{14} we get precisely the same result \rf{21}.
Thus eqs. \rf{13} - \rf{15} give the unique non--singular result
\bea
W_r & = & e^{i \th l} a~r^{- 2+ |l - 1|} + \ldots~~~~, \nn
W_\th & = & e^{i \th l} ~\frac{ia(1 + |l - 1|)}{l - 2} ~
r^{- 2 + |l - 1|} + \ldots~~~~.             \label{22}
\eea

Next let us consider the energy,
\bea
\ve & = & \left| \frac{\pa W_\th}{\pa r} - \frac{il}{r}~
W_r +   \frac{1}{r} ~W_\th - ig Z_\th W_r \right|^2 \nn
&& + ig Z_{r\th} (W_r W^\dg_\th - W_\th W^\dg_r) +\frac{g^2}{2}
|\phi|^2 (W_r W^\dg_r + W_\th W^\dg_\th) \nn
&& - \frac{g^2}{2} (W_r W^\dg_\th - W^\dg_r W_\th)^2 +
\mbox{other terms}~~~~,                        \label{23}
\eea
where we included the kinetic energy of the $W$'s and the magnetic moment terms
\rf{1}.

Consider now the kinetic term.
{}From \rf{22} its behavior is $(r^{p-1})^2$.
However, it is easy to see that
\beq
\frac{\pa W_\th}{\pa r} -  \frac{il}{r} W_r + \frac{1}{r} W_\th = 0
{}~~\mbox{to order}~~r^{p-1}~~~~.                 \label{24}
\eeq
because of \rf{21} and \rf{22}.
Also $|Z_\th W_r|^2 \sim O(r^{2p + 2}) ~,~ |\phi|^2 |W|^2 \sim
O(r^{2p + 2})$.
Thus, to leading order
\bea
\Del & = & ig Z_{r\th} (W_r W^\dg_\th - W_\th W^\dg_r) \nn
& = & 2g a^2 Z_{r\th} (r=0) ~\frac{1 + |l - 1|}{l-2}~
r^{-4 + 2|l-1|}~~~~.                         \label{25}
\eea
For $l = -1, -2, \ldots$~ we see that $\Del < 0$.
Hence, in this case the anomalous magnetic moment effect discussed
previously \cite{6}-\cite{7} is possible.

Let us take $l = -1$.
Then eq. \rf{22} becomes
\bea
W_r & = & e^{- i \th} \left(a + O(r^2)\right)~~~~\nn
W_\th& = & -i e^{- i \th} \left(a + O(r^2)\right)~~~~,  \label{26}
\eea
where the order of the next--to--leading order follows from the
fact that eqs. \rf{13}--\rf{15} are invariant under $r \ra - r$
(using also the equation for $\phi$ and $Z_0$).

At first sight eq. \rf{26} looks peculiar, since $W_r$ and $W_\th$ are
multivalued for $r = 0$.
This is, however, not a problem.
First, let us notice that polar coordinates are ill--defined for $r = 0$.
Therefore the components $W_r$ and $W_\th$ of the vector
${\bf W}$ are defined relative to unit vectors in the $r$-- and
$\th$--directions which are themselves ill--defined for $r = 0$.
Therefore at $r = 0$ it is better to consider ${\bf W}$ decomposed
into $x$-- and $y$--components.
One then finds that $W_x = a \ , \ W_y = - ia$.
The vector ${\bf W}$ is therefore perfectly single--valued for
$x = y = 0$.
For $r \neq 0$ it is, however, more convenient to use $W_r$ and $W_\th$,
so we shall continue to do this with the tacit understanding that if
$r = 0$ one should transform to $x$-- and $y$--coordinates.

For $p=0$ the dominant terms in the energy are thus $\Del$, with $\Del < 0$,
the quartic $(W^4)$ terms and the term $\frac{1}{2} {Z_{r \th}}^2$.
The $r = 0$ energy is minimized for
\beq
2ga^2 = Z_{r \th} (r = 0)~~~~.                        \label{27}
\eeq
When this is the case we also have $F^3_{\mn} \ (r=0) = 0$ and
$D_\mu W_\nu - D_\nu W_\mu = 0 \ (r=0)$.
Due to the fact that $|\phi (r = 0)| = 0$, we see that the full
$SU(2) \otimes U_y(1)$ symmetry is restored along the axis $r=0$,
and the $W$--fields are just pure gauge fields along this axis.
Thus, one can say that the minimum energy $(r = 0) \ W$--dressed
electroweak string interpolates between two different vacua:
The $r = 0 \ SU(2) \otimes U_y(1)$ symmetric vacuum and the
$r = \infty$ broken vacuum.
For this solution the constant $a$ is determined by $Z_{r \th} \ (r=0)$
according to eq. \rf{27}.
When we take $r > 0$ the energy becomes more complicated, with contributions
from the Higgs field also.
A detailed study of the energy can only be done numerically.

The quantity $Z_{r\th} (r=0)$ is in principle determined by the
quantization condition for the $Z$--string,
\beq
\int d^2x Z_{12} (x) = \frac{4 \pi}{g}~~~~,         \label{28}
\eeq
which follows from the asymptotic behavior $Z_\th \ra 2/gr$ ~~for ~$r \ra
\infty$.
A precise determination requires numerical calculations.
However, we have computed $W_r, W_\th, Z_\th$, and $\phi$
in the first two orders in an expansion around $r=0$.
For $l = -1$ we obtain from eqs. \rf{13} -- \rf{15}
\bea
e^{i \th }W_r &= &a +  \frac{1}{2} (g^2a^2 - f - \frac{1}{2} gZ_0)
ar^2 + \ldots~~~~, \nn
e^{i \th} W_\th &=& -ia -  \frac{i}{6}(5 g^2a^2 - f-\frac{5}{2} g Z_0)
ar^2 + \ldots~~~~,                   \label{29}
\eea
where $Z_0 = Z_{r \th} (r =0)$, and where
\beq
|\phi|^2 = \phi'^2_0 r^2 (1 + fr^2)    ~~~~.      \label{30}
\eeq
Again it is remarkable that the {\sl three} eqs. \rf{13}--\rf{15}
produce a consistent answer for the {\sl two} coefficients multiplying
$r^2$ in eqs. \rf{29}.

We have also computed the $Z$--field.
{}From eqs. \rf{9} and \rf{29} we obtain
\bea
Z_\th & = & \frac{1}{2} Z_0 r + \g r^3 + \ldots~~, \nn
\g & = & - \frac{1}{8} g \phi'^2_0 - \frac{1}{24} ga^2
(- 4g^2 a^2 + 8f + 5g Z_0)~~~~.                \label{31}
\eea
Finally we studied the equation for the Higgs field,
\bea
 -  \frac{1}{r}~ \frac{d}{dr} r \frac{d}{dr}
|\phi|& + &\left(\frac{1}{r} - \frac{g}{2}
{}~Z_\th \right)^2 |\phi| - 2 \lam \phi^2_0 |\phi| + \nn
& + & \frac{1}{2} g^2 W_i W^\dg_i |\phi| + 2 \lam |\phi|^3 = 0~~~~.
                                    \label{32}
\eea
The constant $f$ in eq. \rf{30} is given by
\beq
f = - \frac{1}{9} gZ_0 -\frac{4}{9} (\lam \phi^2_0 - \frac{1}{2} g^2 a^2)
{}~~~~.                              \label{33}
\eeq

{}From eq. \rf{31} we see that to order $r^3$ the effect of $W$--condensation
is to increase $\g$ by the quantity $+ \frac{1}{6} g^3 a^4$.
Thus the $Z$--field is enhanced. This is an effect of the anti--screening
property of the electroweak vacuum discussed in ref. \rf{7}.

For $r \ra \infty$ the asymptotic behaviors are
\bea
W_r, W_\th & \sim  & e^{-m{_v}r}~~~,~~~ m^2_v = \frac{1}{2} g^2 \phi^2_0~~~,
                                         \nn
\phi_0 - |\phi| &\sim & e^{- m{_H}r}~~~,       \nn
Z_\th - \frac{2}{gr} & \sim & e^{- m_{v}r} ~~~~. \label{34}
\eea
Due to eqs. \rf{32} and \rf{34} the $W$'s have very little influence
on $\phi_0 - |\phi| $ for $r \ra \infty$.

The conclusion is thus that it appears likely that there exists a
$W$--dressed electroweak string for $\th_W = 0$, the consistency
of which is closely related to the integrability condition \rf{15}.
In principle, one should study the stability of this object by
considering small perturbations.
However, we think it is likely that the string is metastable, since
the phenomenon of $W$--condensation originally was found by a stability
analysis \cite{9} showing that it pays to have $W$--fields for large
external fields, and since a similar analysis was later performed
explicitly for the electroweak string by Perkins \cite{6}.

This does, of course, not imply that the $W$--dressed electroweak string
is absolutely stable.
For example, adding Chern--Simons terms Poppitz has shown \cite{10}
that $W$--condensation leads to a string with a periodic structure
in the $z$--direction if one has a finite fermion density.
Such a string could also lead to baryon asymmetry.

  We mention that even
if one has more than one Higgs multiplet \cite{5}, it is in any case
possible to lower the energy by $W$--condensation, as long as one
has terms of the type \rf{1} in the energy.

Finally we mention that one can see from the equations of motion
that for $\th_W \neq 0$ a magnetic field $F_{12} = \pa_1 A_2 - \pa_2 A_1$
is necessary in the string core if a $W$--dressed $Z$--string exists.

I thank Tanmay Vachaspati for many interesting discussions.

\end{document}